\documentclass[sigplan, screen, nonacm]{acmart}

\setcopyright{none}           
\renewcommand\footnotetextcopyrightpermission[1]{} 

\acmConference[]{}{}{}
\acmYear{}
\acmMonth{}

\AtBeginDocument{%
  }

\begin{document}

\title{Stencil Computations on Tenstorrent Wormhole}

\author{Lorenzo Piarulli}
\email{piarulli@di.uniroma1.it}
\orcid{0009-0006-8071-8806}
\affiliation{%
  \institution{Sapienza University of Rome}
  \country{Rome, Italy}
}
\author{Daniele De Sensi}
\email{desensi@di.uniroma1.it}
\orcid{0000-0002-7244-639X}
\affiliation{%
  \institution{Sapienza University of Rome}
  \country{Rome, Italy}
}

\renewcommand{\shortauthors}{Piarulli et al.}

\begin{abstract}
As investment in AI-focused accelerators grows and their deployment in supercomputing facilities expands, understanding whether these architectures can efficiently support traditional scientific kernels is critical for the future of High-Performance Computing. We investigate the mapping of 2D 5-point stencil computations onto the Tenstorrent Wormhole, a RISC-V AI dataflow accelerator. We develop two heterogeneous implementations: \textit{Axpy}, which decomposes the stencil into element-wise submatrix operations, and \textit{MatMul}, which reformulates it as a matrix multiplication. While the CPU baseline remains $3\times$ faster end-to-end, profiling reveals that the isolated Wormhole kernel is competitive with CPU execution, with the gap driven by PCIe transfers, device initialization, and host-side preprocessing. Despite slower runtime, \textit{Axpy} achieves lower energy consumption than the CPU baseline for large inputs. Through detailed profiling and theoretical analysis, we identify key architectural and software limitations of the current platform and outline concrete hardware and software directions that could make AI accelerators competitive for HPC workloads.
\end{abstract}

\maketitle

\section{Introduction}
\label{sec:intro}

High-Performance Computing (HPC) methods are fundamental for scientific discovery, from weather forecasting and climate modeling to computational genomics and materials science~\cite{zeni2020logan, giannozzi2020quantum}.
Many of these scientific challenges are modeled using Partial Differential Equations (PDEs), which require iterative approximation algorithms that are computationally intensive, frequently demanding large-scale computations running for days or weeks on supercomputers.

At the same time, the explosive growth of AI workloads has redirected supercomputing resources toward deep learning. This trend is driving substantial investment in domain-specific accelerators (DSAs) optimized for matrix-heavy AI tasks~\cite{schmidhuber2015deep}, and it is increasingly likely that future HPC systems will dedicate large partitions to these accelerators. This raises a critical question: can these AI-focused architectures also serve traditional HPC kernels such as stencil computations, or will the scientific computing community be left behind?

Tenstorrent Wormhole~\cite{tt_wormhole_specs} exemplifies this new class of accelerators.
It is a RISC-V-based design built around a grid of Tensix cores, each executing dataflow operations on fixed 32$\times$32 tiles, primarily designed for GEMM-based AI workloads.
Unlike GPUs with mature kernel-level ecosystems (CUDA, ROCm), Wormhole represents an evolving architecture with a kernel-level programming model (TT-Metalium) that gives developers direct control over hardware resources.

In this paper, building on Brown et al.'s work on Tenstorrent Grayskull~\cite{brown2024stencils}, we develop two heterogeneous CPU--Wormhole stencil implementations and conduct an extensive analysis of their performance, energy efficiency, and architectural bottlenecks.

Our approach adopts a heterogeneous execution model that performs frequent data transfers and CPU-side format conversions at every iteration, diverging from conventional accelerator best practices. This choice stems from a practical constraint: Brown and Barton's work on Grayskull~\cite{brown2024stencils} demonstrated that performing scalar operations entirely on-device is prohibitively expensive due to the Tensix cores' lack of efficient scalar support, and we were unable to find an alternative on-device solution for Wormhole. We therefore opted for a CPU--accelerator work division, accepting the performance penalty of repeated transfers in exchange for a working implementation that could reveal architectural and API limitations and suggest concrete directions for future hardware and software improvements. In addition, this model also evaluates unified memory architectures, such as AMD MI300A~\cite{amd_mi300a} or NVIDIA GH200~\cite{nvidia_gh200}, where transfer penalties disappear and hybrid execution scheme becomes more efficient.

Our contributions are as follows:
\begin{enumerate}
\item Two heterogeneous methodologies which we denote as \textit{Axpy} and \textit{MatMul}, for mapping 2D 5-point stencil computations onto Wormhole, with a clear analysis of each method's limitations.
\item A detailed characterization of Wormhole's limitations through profiling with Tracy, identifying bottlenecks on both the hardware and software side and proposing architectural and API improvements to enable efficient HPC computation on future Tenstorrent accelerators.
\item A theoretical analysis of Unified Virtual Memory (UVM) and Unified Physical Memory (UPM) scenarios showing that unified memory could make the heterogeneous approach competitive, suggesting this as a promising architectural direction. We further discuss how Tenstorrent's Blackhole architecture could address the identified limitations for future HPC workloads on Tenstorrent hardware.
\end{enumerate}

\section{Related Work}
\label{sec:related}

\textbf{HPC on Tenstorrent accelerators.}
Brown and Barton~\cite{brown2024stencils} presented the first systematic evaluation of stencil computations on the Tenstorrent Grayskull architecture.
Their \textit{Axpy}-style approach demonstrated higher energy efficiency than conventional CPUs but was limited by Grayskull's LPDDR memory bandwidth and the inefficiency of scalar operations on the lightweight Tensix RISC-V cores.
More recently, Brown et al.~\cite{brown2025fft} ported the Cooley--Tukey FFT algorithm to Wormhole, demonstrating that while the accelerator is slower than a 24-core Xeon Platinum CPU, it draws approximately 8$\times$ less power and consumes 2.8$\times$ less energy.
Amati et al.~\cite{amati2025nbody} accelerated gravitational N-body simulations on the Wormhole n300, achieving more than 2$\times$ speedup and approximately 2$\times$ energy savings compared to optimized CPU implementations.
These works collectively indicate that Tenstorrent accelerators, while not yet performance-competitive for all HPC workloads, offer compelling energy efficiency advantages.

\textbf{Stencils on tensor cores.}
Chen et al.~\cite{chen2024convstencil} proposed ConvStencil, which transforms stencil computations into matrix multiplications to leverage GPU tensor cores.
Their approach demonstrated significant speedups on NVIDIA GPUs by reformulating the stencil as an im2col-style transformation followed by GEMM.
Our \textit{MatMul} method draws inspiration from this work but adapts it to Wormhole's tiled execution model, which imposes stricter alignment constraints than GPU tensor cores.

\textbf{Architectural characterization.}
Cai et al.~\cite{cai2023assessing} assessed Tenstorrent's matrix multiply acceleration capabilities, providing early insights into the architecture's strengths and limitations.
A recent microbenchmarking study~\cite{blackhole_microbench} dissected the Blackhole architecture empirically, confirming that the Tensix dataflow model achieves high utilization of internal bandwidth when computation and communication are properly overlapped.

\section{Background}
\label{sec:background}

\subsection{Stencil Computations}

Stencil computations are foundational HPC kernels used to solve PDEs through finite-difference methods (FDM).
In this work, we solve the 2D Laplace equation $\Delta u = 0$ using the Jacobi iterative method.
The discretized update rule for the 5-point stencil is:
\begin{equation}
  u^{(k+1)}_{i,j} = \frac{1}{4}\left(u^{(k)}_{i+1,j} + u^{(k)}_{i-1,j} + u^{(k)}_{i,j+1} + u^{(k)}_{i,j-1}\right)
  \label{eq:stencil}
\end{equation}
where $(i,j)$ denotes the grid position and $k$ the iteration index. This formulation applies Dirichlet boundary conditions (zero-valued boundaries) and iterates for a fixed number of iterations rather than until convergence.
Although conceptually simple, efficiently mapping this pattern onto tile-based architectures requires careful data layout management since the stencil accesses neighboring elements that may span different tiles.

\subsection{Tenstorrent Wormhole Architecture}

\begin{figure}[t]
  \centering
  \includegraphics[width=0.7\columnwidth]{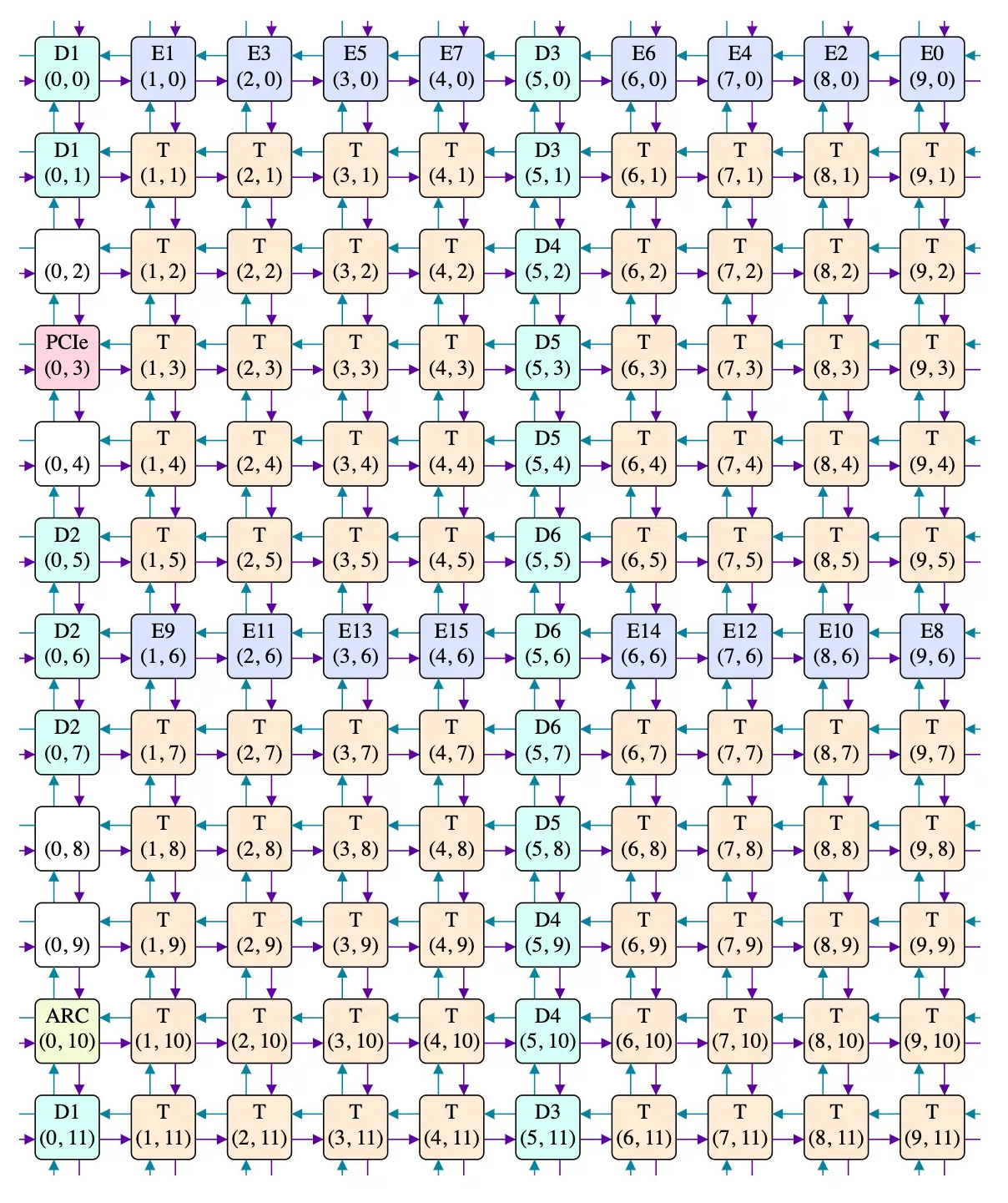}
  \caption{Tenstorrent Wormhole architecture. The 10$\times$12 grid contains 64 Tensix cores (T), DRAM controllers (D), Ethernet interfaces (E), PCIe controller, and ARC management core (image from \cite{Corsix2024}).}
  \label{fig:wormhole_arch}
\end{figure}

Tenstorrent Wormhole is a RISC-V AI accelerator available as a PCIe card (n150d/n300d variants).
The architecture, shown in Figure~\ref{fig:wormhole_arch}, is built around a mesh of 64 Tensix cores interconnected through a high-bandwidth Network-on-Chip (NoC), complemented by DRAM banks and Ethernet interfaces for multi-accelerator scaling.

\begin{figure}[t]
  \centering
  \includegraphics[width=\columnwidth]{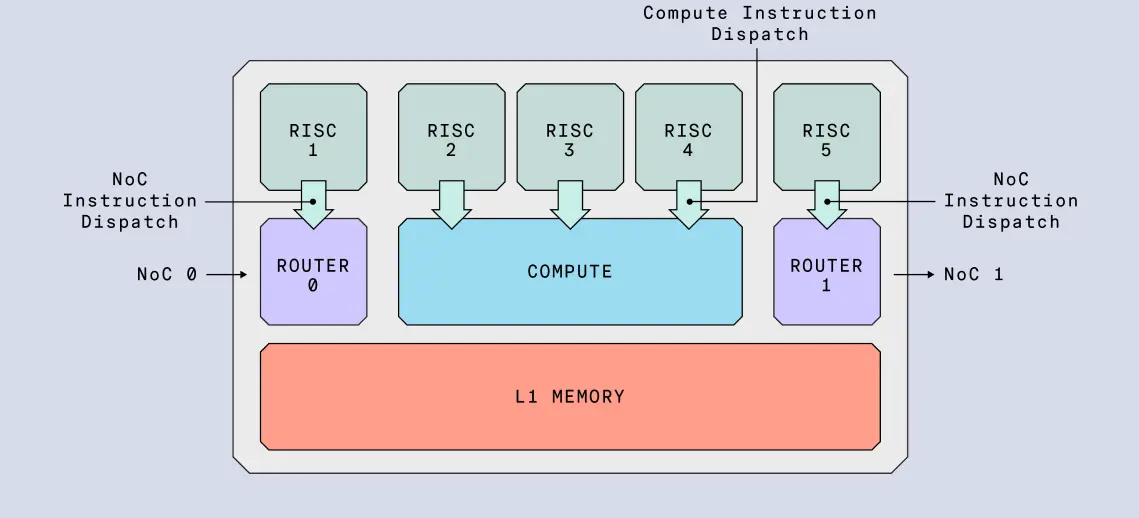}
  \caption{Tensix core architecture. Five baby RISC-V cores coordinate NoC data movement (Router 0/1), computation (Compute block with matrix and vector engines), and local L1 memory access (image from \cite{TenstorrentTensixNeo}).}
  \label{fig:tensix_core}
\end{figure}

Each Tensix core (Figure~\ref{fig:tensix_core}) is a self-contained dataflow processor integrating five lightweight ``baby'' RISC-V CPUs, a matrix engine (FPU) optimized for 32$\times$32 tile operations, a vector/SIMD unit (SFPU), pack/unpack units for data format conversion, two NoC interfaces, and approximately 1.5\,MB of local SRAM.
The five RISC-V cores serve distinct roles: two handle data movement (NoC reader/writer), while three manage computation through a pipelined Unpack$\rightarrow$Math$\rightarrow$Pack paradigm.
The cores synchronize through a shared destination register (DST) using explicit lock/release primitives.

A critical constraint for HPC workloads is that Wormhole operates natively on 32$\times$32 tiles.
Although smaller tile sizes are theoretically supported, key API functions such as \texttt{tilize\_nfaces()} and \texttt{untilize\_nfaces()} currently operate only on 32$\times$32 tiles.
Furthermore, all data transferred to the accelerator must be converted from row-major layout to a tiled memory layout (\texttt{tilize}), and results must be converted back (\texttt{untilize}).
The TT-Metalium programming model provides kernel-level access to hardware resources through C++ APIs compiled Just-In-Time at runtime.
All experiments in this work use TT-Metal version 0.62.2 with \emph{bfloat16} precision.

Table~\ref{tab:specs} summarizes the key specifications of the Wormhole n150d used in our experiments.

\begin{table}[t]
  \centering
  \caption{Tenstorrent Wormhole n150d specifications.}
  \label{tab:specs}
  \small
  \begin{tabular}{@{}ll@{}}
    \toprule
    \textbf{Specification} & \textbf{Value} \\
    \midrule
    Tensix Cores        & 72 (64 usable)\\
    Clock            & 1\,GHz \\
    SRAM                & 108\,MB \\
    Memory              & 12\,GB GDDR6 \\
    Memory Bandwidth    & 288\,GB/s \\
    TeraFLOPS (FP8)     & 262 \\
    TeraFLOPS (FP16)    & 74 \\
    Total Board Power   & 160\,W \\
    System Interface    & PCIe Gen4 $\times$16 \\
    \bottomrule
  \end{tabular}
\end{table}
\section{Methodology}
\label{sec:methodology}

\subsection{Heterogeneous Execution Model}

Both of our methodologies follow a heterogeneous CPU--accelerator execution model that deliberately diverges from conventional accelerator usage. Typically, data is transferred to an accelerator once, all iterations execute on-device, and only the final result is transferred back. However, this model is not always feasible due to architectural or software limitations. Consequently, results must be transferred back to the CPU after each iteration for scalar preprocessing; this introduces significant data movement overhead before they are sent back for the subsequent iteration.

This design choice is motivated by one factor: \textbf{Wormhole's inability to efficiently handle scalar operations}. The architecture excels at tile-based pipelined MIMD operations but is highly inefficient for scalar and irregular computations.
The baby RISC-V cores within each Tensix are lightweight processors not optimized for intensive scalar work~\cite{brown2024stencils}.
Performing data extraction and boundary handling on the accelerator, as attempted by Brown et al. on Grayskull~\cite{brown2024stencils}, proved extremely expensive due to the architecture's poor scalar performance and strict memory alignment constraints.
Conversely, CPUs are optimized for scalar manipulations, cache-friendly data movement, and control-heavy operations.

Both implementations are therefore structured in two phases per iteration: data \textit{preprocessing} and \textit{computation}. In the \textit{Axpy} method (Sec.~\ref{sec:axpy}), the CPU extracts the shifted submatrices (\textit{data preprocessing}), while the Tenstorrent accelerator computes the element-wise sum and scaling (\textit{computation}). In the \textit{MatMul} method (Sec.~\ref{sec:matmul}), the CPU converts the input to stencil-to-row format (\textit{data preprocessing}), and the Tenstorrent accelerator performs the matrix multiplication (\textit{computation}). At the end of each iteration, the result is transferred back to the CPU before the next iteration begins.

\subsection{\textit{Axpy} Method}
\label{sec:axpy}

Inspired by Brown's Grayskull work~\cite{brown2024stencils}, we decompose the 5-point stencil into submatrix operations (Figure~\ref{fig:axpy_pipeline}).

The method proceeds as follows for each iteration:

\begin{figure}[t]
  \centering
  \includegraphics[width=\columnwidth]{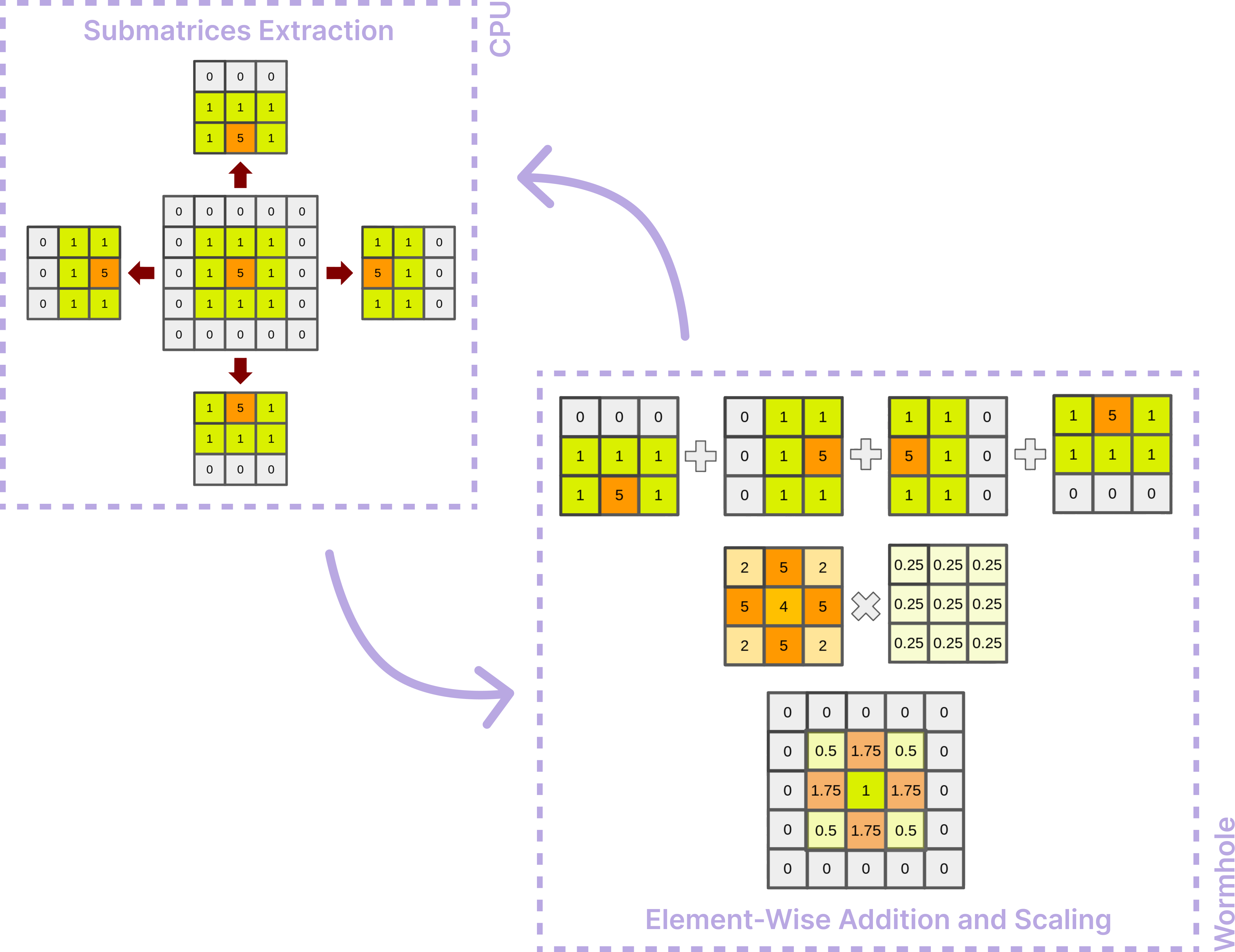}
  \caption{\textit{Axpy} pipeline. From the padded input the submatrices are extracted, the submatrices data are sent to Wormhole for element-wise addition and scaling.}
  \label{fig:axpy_pipeline}
\end{figure}

\textbf{CPU phase.}
From the padded input matrix, extract four shifted submatrices corresponding to the stencil neighbors: up (shift one row upward), down (shift one row downward), left (shift one column left), and right (shift one column right).
Each submatrix is stored in a contiguous buffer and padded so that its total element count is divisible by $32 \times 32 = 1024$, satisfying Wormhole's tile alignment.
In the first iteration, the input is padded with Dirichlet boundary conditions (halo of zeros).
In subsequent iterations, padding and extraction are fused into a single optimized routine.

\textbf{Wormhole phase.}
The four submatrix buffers are transferred over PCIe to Wormhole.
Tiles are distributed across all 64 Tensix cores: for an $N \times N$ matrix with tile size 32, the total number of tiles is $T = (N/32)^2$, and each core receives $\lceil T/64 \rceil$ tiles.
For each tile index $k$, the compute kernel executes:
\begin{equation}
\footnotesize
  \text{out}^{32\times32}_{k} = 0.25^{32\times32} \odot \left(\text{up}^{32\times32}_{k} + \text{down}^{32\times32}_{k} + \text{left}^{32\times32}_{k} + \text{right}^{32\times32}_{k}\right)
  \label{eq:axpy_kernel}
\end{equation}
using the matrix engine for element-wise addition and element-wise product (denoted as $\odot$). The scalar 0.25 is represented as a constant 32$\times$32 tile for compatibility with the matrix engine.
Results are written back to device DRAM and then transferred to CPU DRAM over PCIe.

The key advantage of \textit{Axpy} is that data remains in row-major layout throughout; no \texttt{tilize}/\texttt{untilize} conversion is needed because element-wise addition is layout-agnostic.
The only requirement is that buffer sizes are multiples of the tile element count.

\textbf{Limitations.}
The fixed 32$\times$32 tile size mandates padding for arbitrary input dimensions, wasting memory and compute cycles on padding elements.
Furthermore, the per-iteration CPU--device round-trip over PCIe Gen4 ($\approx$31.5\,GB/s per direction) introduces transfer overhead that dominates.

\subsection{\textit{MatMul} Method}
\label{sec:matmul}

\begin{figure}[t]
  \centering
  \includegraphics[width=\columnwidth]{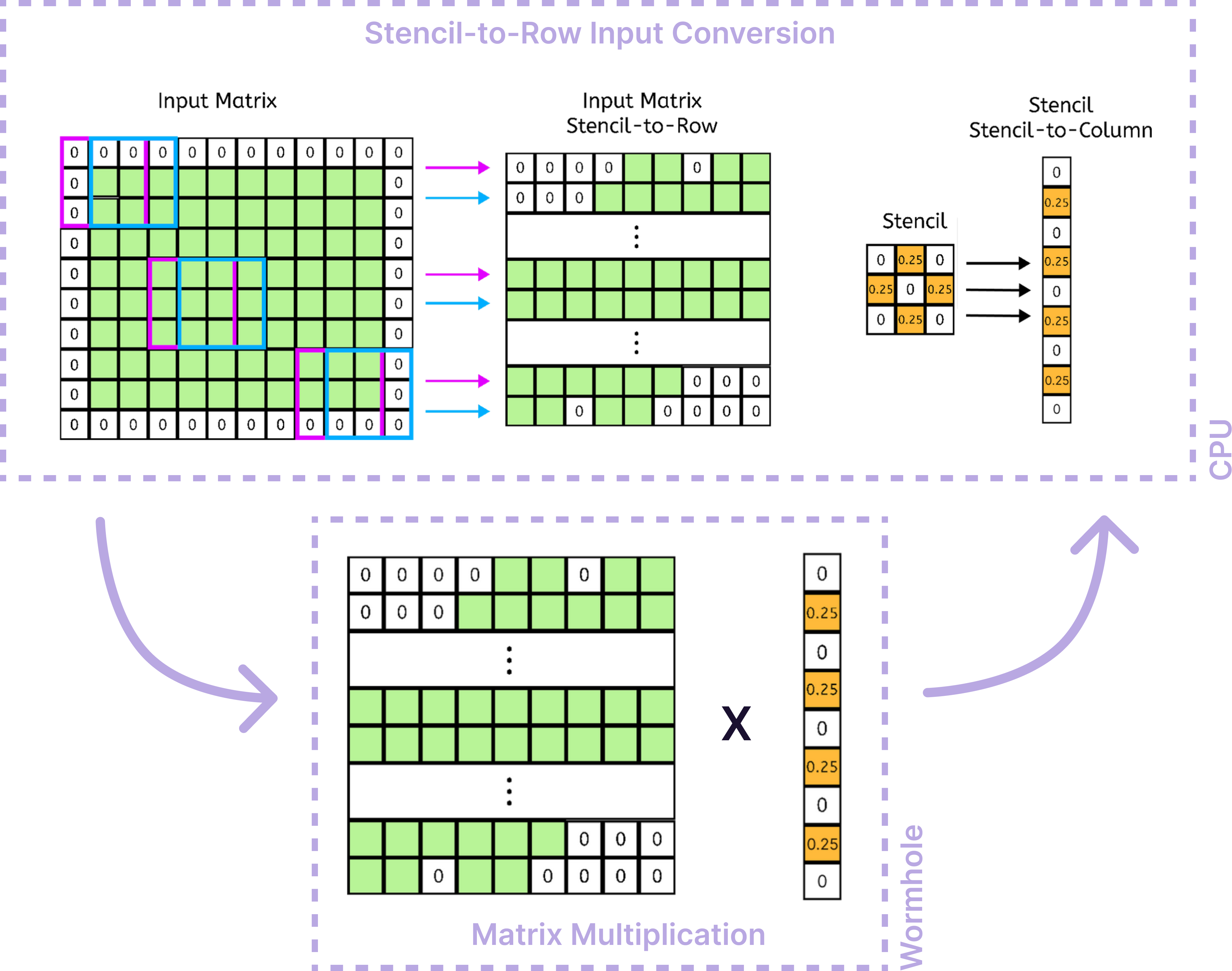}
  \caption{\textit{MatMul} pipeline. The padded input is converted to stencil-to-row format, the stencil kernel is flattened to a column vector, both are aligned to 32$\times$32 tiles via padding, and the tilized data is sent to Wormhole for matrix multiplication.}
  \label{fig:matmul_pipeline}
\end{figure}

In our second methodology, inspired by ConvStencil~\cite{chen2024convstencil}, we reformulate the stencil as a matrix multiplication (Figure~\ref{fig:matmul_pipeline}).
This leverages Wormhole's highly optimized matrix multiplication engine.

\textbf{CPU phase.}
The padded input is converted to stencil-to-row format (denoted as $In$): for each grid point, the $3\times3$ neighborhood is unrolled into a row of 9 elements.
An $N \times N$ grid produces an $(N^2) \times 9$ matrix.
The stencil weights are flattened into a $9 \times 1$ column vector (denoted as $St$).
Both matrices must then be aligned to 32$\times$32 tiles by padding: the column vector is padded to $32 \times 1$ and replicated 32 times to form a $32 \times 32$ tile; the input matrix rows are padded from 9 to 32 columns.
Finally, \texttt{tilize\_nfaces()} converts the row-major data to Wormhole's tiled memory layout.

\textbf{Wormhole phase.}
Tiles are distributed across Tensix cores and tiled matrix multiplication is executed, using the batched matrix multiplication paradigm.
For each tile index $k$: 
\begin{equation}
  \text{out}^{32\times32}_{k} = \text{In}^{32\times32}_{k}\text{St}^{32\times32}
  \label{eq:axpy_kernel}
\end{equation}

After completion, \texttt{untilize\_nfaces()} converts results back to row-major layout, and the CPU extracts the final grid for the next iteration.

\textbf{Limitations.}
The \textit{MatMul} method incurs severe overhead from three sources.
First, the stencil-to-row transformation expands an $8 \times 8$ matrix (128\,B in FP16) to 4096\,B after tiling, a 32$\times$ memory increase.
Second, the \texttt{tilize}/\texttt{untilize} conversions are performed by CPU-side utility functions that account for approximately 90\% of total CPU-side execution time.
Third, the larger memory footprint limits the maximum feasible input size: while \textit{Axpy} handles up to $30720 \times 30720$, \textit{MatMul} saturates DRAM at $16384 \times 16384$.
\section{Experimental Results}
\label{sec:results}

\subsection{Experimental Setup}

Experiments were conducted on a node of the Sapienza Computer Science Department cluster equipped with two AMD EPYC 7301 16-core processors (64 virtual cores total), 256\,GB DRAM, and a Tenstorrent Wormhole n150d accelerator.
All code was compiled with Clang 17 at \texttt{-O3} with \texttt{-march=native}.
Problem sizes range from $1024^2$ to $30720^2$ elements in bfloat16 precision, with iteration counts of 100, 500, and 1000.
The CPU baseline is an OpenMP-parallelized stencil implementation with SIMD optimizations enabled via compiler flags.

\begin{figure}[t]
  \centering
  \includegraphics[width=\columnwidth]{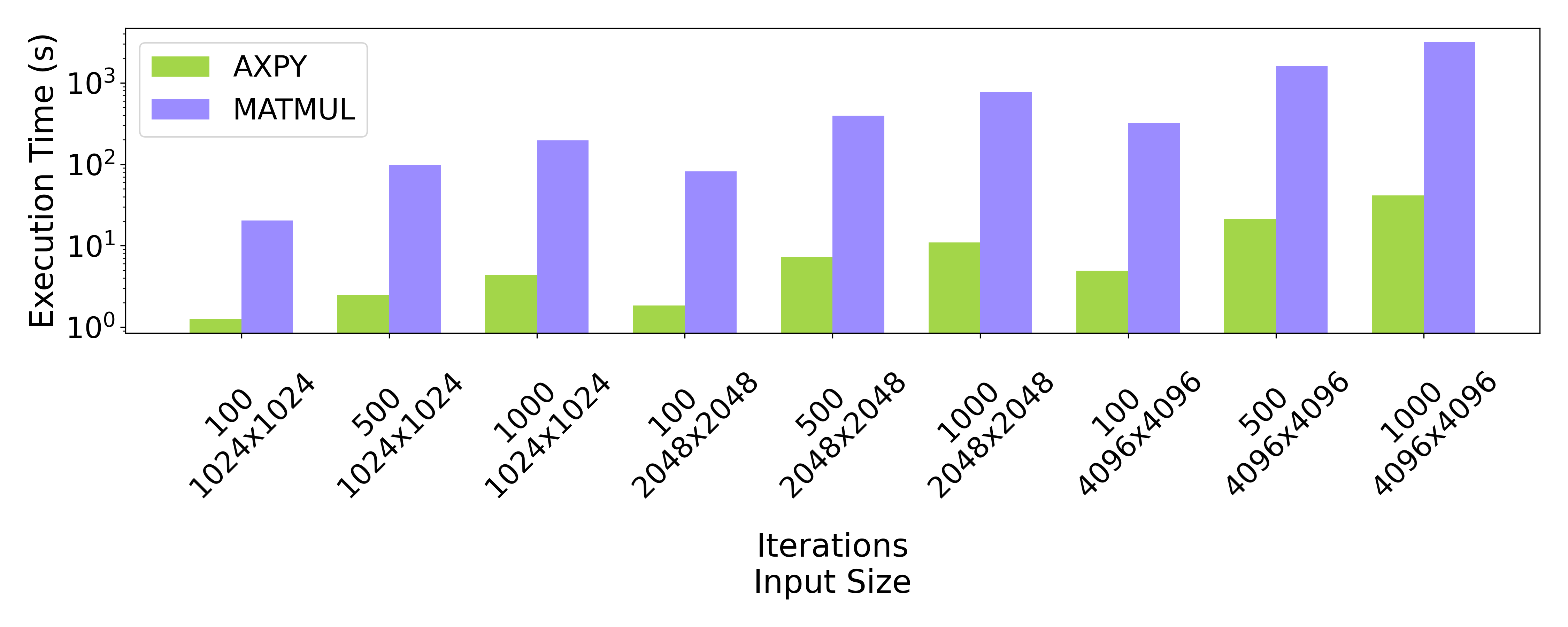}
  \caption{Execution time comparison between \textit{Axpy} and \textit{MatMul} heterogeneous implementations (log scale). \textit{Axpy} is approximately 75$\times$ faster across all configurations.}
  \label{fig:axpy_vs_matmul}
\end{figure}

\subsection{\textit{Axpy} vs.\ \textit{MatMul} Comparison}

Figure~\ref{fig:axpy_vs_matmul} compares the two heterogeneous implementations.
\textit{Axpy} outperforms \textit{MatMul} by approximately 75$\times$ across all tested configurations.
The gap widens with input size, consistent with the quadratic growth of the stencil-to-row expansion in \textit{MatMul}.

\begin{figure}[t]
  \centering
  \includegraphics[width=\columnwidth]{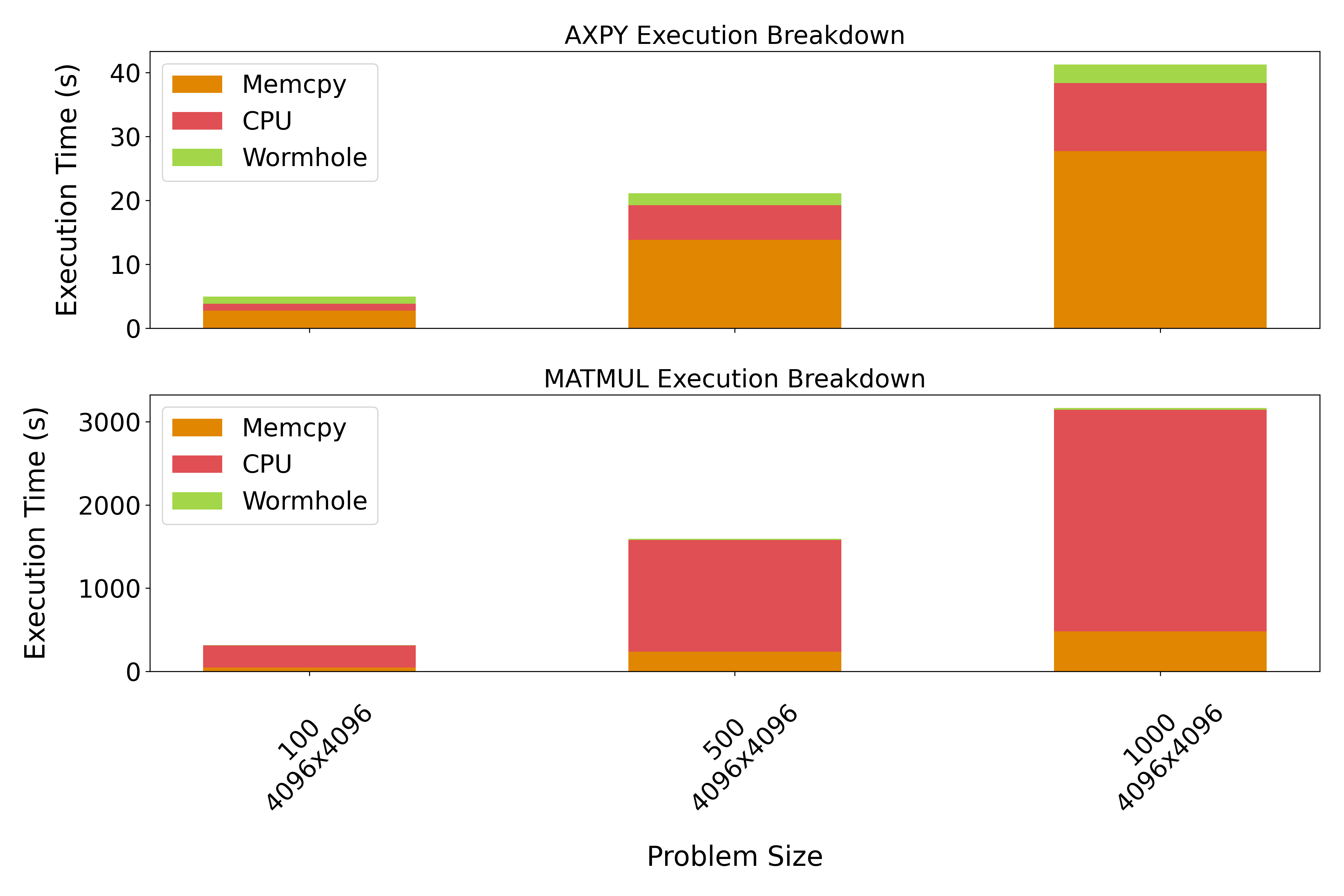}
  \caption{Execution time breakdown by phase. Top: \textit{Axpy} shows balanced distribution across CPU, memcpy, and Wormhole. Bottom: \textit{MatMul} is dominated by CPU-side tiling conversions ($\approx$90\%).}
  \label{fig:phase_breakdown}
\end{figure}

The phase breakdown in Figure~\ref{fig:phase_breakdown} reveals the root cause.
In \textit{Axpy}, execution time is distributed across CPU tasks (15\%), memory copies (25\%), and Wormhole computation (60\%), indicating a reasonable balance.
In contrast, \textit{MatMul} spends approximately 90\% of its time on CPU-side work---dominated by the \texttt{tilize\_nfaces()} and \texttt{untilize\_nfaces()} functions---with only 5\% each for memory copies and Wormhole computation.
The Wormhole engine is thus largely idle in the \textit{MatMul} variant, negating the advantage of using optimized matrix multiplication hardware.

\subsection{\textit{Axpy} vs.\ CPU Baseline}

\begin{figure}[t]
  \centering
  \includegraphics[width=\columnwidth]{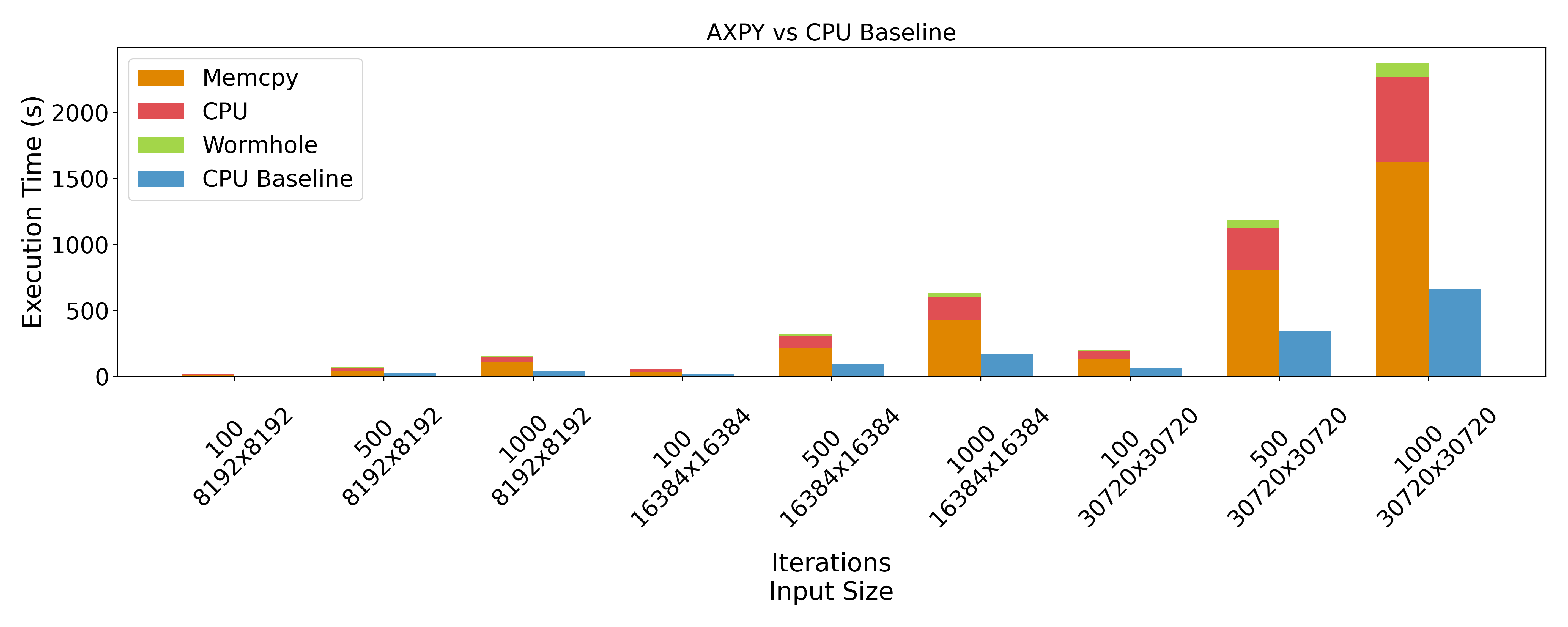}
  \caption{Execution time comparison between \textit{Axpy} and CPU baseline implementations.}
  \label{fig:axpy_vs_cpu}
\end{figure}

Given \textit{Axpy}'s clear superiority, we evaluate it against the optimized CPU baseline.
The CPU baseline remains approximately 3$\times$ faster than the heterogeneous \textit{Axpy} implementation (Figure~\ref{fig:axpy_vs_cpu}).
This gap is primarily attributed to two factors: PCIe transfer overhead and CPU-side submatrix extraction repeated at every iteration (Figure~\ref{fig:axpy_vs_cpu}).

However, profiling with the Tracy profiler~\cite{tracy} reveals a critical insight.
When we isolate the actual kernel execution time on Wormhole (excluding device initialization, kernel launch, and transfer overhead), the Wormhole kernel is \emph{competitive} with the CPU baseline for the \textit{Axpy} variant.
Table~\ref{tab:kernel_times} quantifies this: the host-observed execution time includes a near-constant overhead of approximately 1s from device initialization that does not scale with input size.
For large inputs ($1024^2$ at 1000 iterations), the pure kernel time is only 124\,ms, while the host-observed time is 1376\,ms---an overhead factor exceeding 10$\times$.

\begin{table}[t]
  \centering
  \caption{Kernel execution time: isolated (Tracy-profiled) vs.\ host-observed (including initialization and transfer overhead).}
  \label{tab:kernel_times}
  \small
  \begin{tabular}{@{}lrr@{}}
    \toprule
    \textbf{Config (iter-input)} & \textbf{Kernel Time} & \textbf{Total Time} \\
    \midrule
    100 -- $128^2$ (\textit{Axpy})    & 0.50\,ms   & 1006\,ms \\
    1000 -- $128^2$ (\textit{Axpy})   & 4.96\,ms   & 1140\,ms \\
    100 -- $1024^2$ (\textit{Axpy})   & 12.6\,ms   & 981\,ms  \\
    1000 -- $1024^2$ (\textit{Axpy})  & 124\,ms    & 1376\,ms \\
    \midrule
    100 -- $128^2$ (\textit{MatMul})  & 2.58\,ms   & 1013\,ms \\
    1000 -- $1024^2$ (\textit{MatMul})& 1358\,ms   & 2460\,ms \\
    \bottomrule
  \end{tabular}
\end{table}

The Tracy-profiled pipelining analysis further confirms that the dataflow paradigm within each Tensix core functions as designed: the Unpack thread prepares data, the Math thread executes arithmetic, and the Pack thread stores results, with effective overlap across tiles.
The Reader and Writer threads stream data concurrently, and all cores exhibit similar timing patterns due to balanced workload distribution and a shared 1\,GHz clock domain.
The dominant bottleneck is not on-chip computation but rather the surrounding infrastructure: device initialization, PCIe transfers, and CPU-side preprocessing.

\subsection{Energy Efficiency}

Energy efficiency is a compelling advantage of the Wormhole architecture.
Using TT-SMI for Wormhole power measurements and the AMD EPYC 7301's TDP (170\,W) as a CPU estimate, we computed energy as $E = \text{Runtime} \times \text{Power}~(P)$.
Wormhole's idle power is approximately 11\,W, rising to 20--24\,W during computation.
Despite being 3$\times$ slower, the heterogeneous \textit{Axpy} implementation consumes \emph{less total energy} than the CPU baseline for large inputs at high iteration counts, if we remove the data movement energy consumption.
This advantage stems from Wormhole's architectural characteristics: high core count for balanced workload distribution and the modest 1\,GHz operating frequency.

When isolating only the Wormhole kernel execution (excluding CPU preprocessing), the energy advantage is even more pronounced: the kernel operates at 20--24\,W for milliseconds, whereas the CPU baseline sustains 170\,W for comparable or longer durations.
This finding suggests that with architectural improvements to reduce transfer and conversion overhead, Wormhole-class accelerators could be highly energy-efficient HPC platforms.

\section{Discussion}
\label{sec:discussion}

\subsection{Identified Limitations}

Our analysis reveals four critical architectural and software limitations of the current Wormhole platform for HPC workloads:

\textbf{Fixed 32$\times$32 tiling constraint.}
All buffers must be aligned to 32$\times$32 tiles, precluding fine-grained computation.
PDE solvers frequently require small vector operations for boundary handling and neighbor access.
Although the hardware is capable of using smaller tiles, the API (\texttt{tilize\_nfaces()}, \texttt{untilize\_nfaces()}) does not support them.
Enabling flexible tile sizes would significantly reduce padding overhead and memory waste.

\textbf{Inefficient scalar and small-vector support.}
Wormhole's baby RISC-V cores are not designed for intensive scalar workloads.
This forces format conversions, boundary handling, and data extraction to be offloaded to the host CPU, creating an inherent bottleneck.
The \textit{MatMul} variant's 90\% CPU time is a direct consequence: the stencil-to-row transformation and tiling conversion are scalar-heavy operations that Wormhole cannot execute efficiently.

\textbf{Tilize/untilize overhead.}
The CPU-side \texttt{tilize\_nfaces()} and \texttt{untilize\_nfaces()} functions introduce prohibitive latency for any methodology requiring tiled data layout.
The \textit{Axpy} method avoids this entirely because element-wise operations are layout-agnostic, but any approach leveraging Wormhole's matrix multiplication engine must pay this cost.
Hardware support for flexible memory layouts, or on-chip tiling engines, would be transformative.

\textbf{PCIe Gen4 bottleneck.}
The discrete CPU--accelerator design imposes $\approx$31.5\,GB/s per-direction bandwidth through PCIe Gen4 $\times$16.
For our iterative methodology, this cost is paid every iteration, dominating total execution time for moderate input sizes.
Tighter CPU--accelerator integration through high-bandwidth interconnects or unified memory would dramatically reduce this overhead.

\subsection{Unified Memory Analysis}

\begin{figure}[t]
  \centering
  \includegraphics[width=\columnwidth]{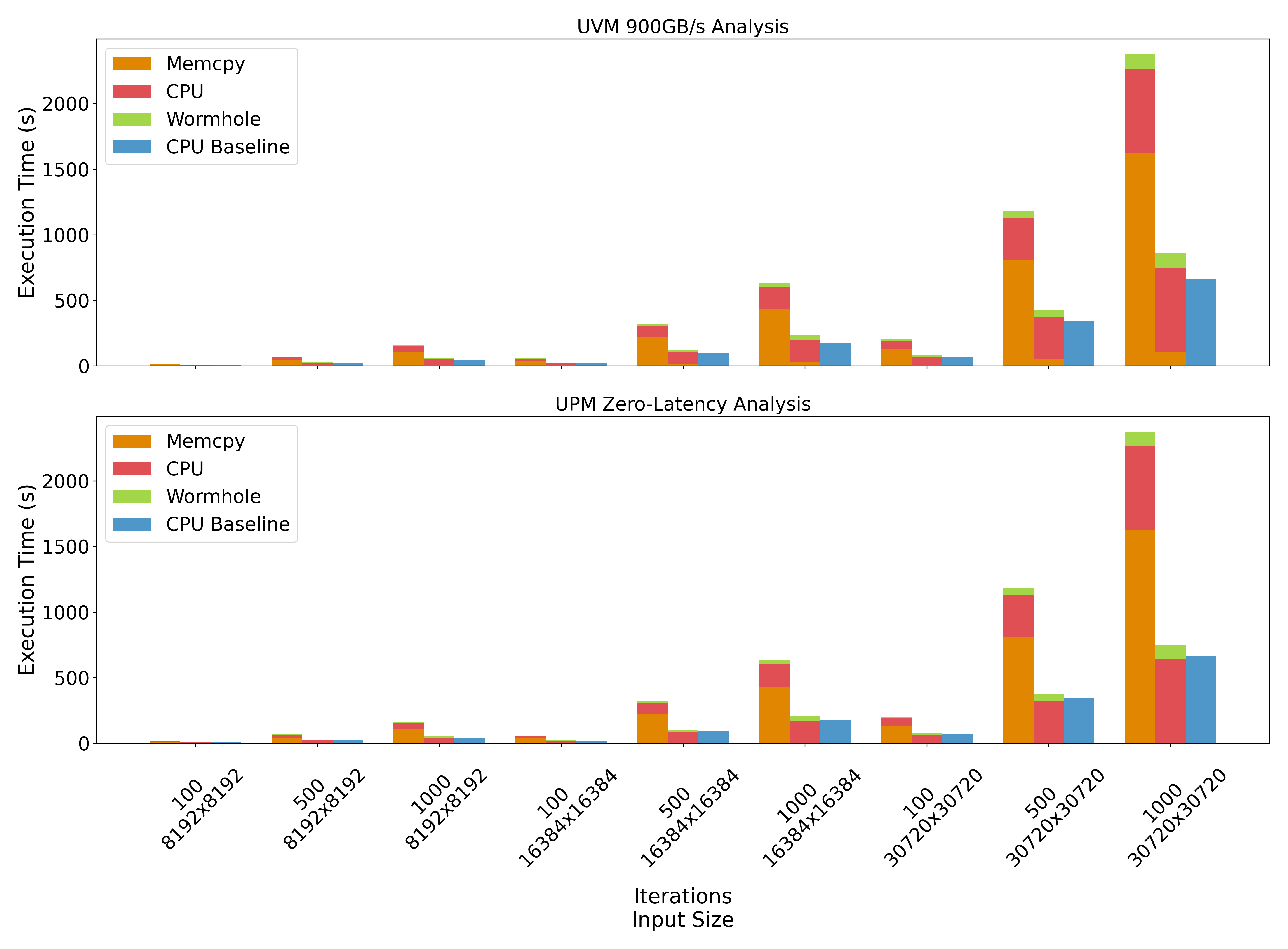}
  \caption{Theoretical performance under UVM (top, 900\,GB/s NVLink-C2C bandwidth with 450\,GB/s per-direction) and UPM (bottom, zero transfer overhead) scenarios compared to the CPU baseline.}
  \label{fig:uvm_upm}
\end{figure}

To quantify the potential of future architectures, we model two unified memory scenarios (Figure~\ref{fig:uvm_upm}).

\textbf{UVM model.}
We replace PCIe Gen4 bandwidth ($\approx$31.5\,GB/s) with NVLink-C2C-class bandwidth ($\approx$450\,GB/s), as found in the NVIDIA GH200 Grace Hopper Superchip~\cite{nvidia_gh200}.
This reduces transfer overhead by approximately 15$\times$.
Under this model, the \textit{Axpy} method approaches CPU baseline performance while maintaining significantly lower energy consumption.

\textbf{UPM model.}
We model complete memory unification, as in AMD MI300A~\cite{amd_mi300a}, where CPU and accelerator share coherent physical memory.
This eliminates transfers entirely and also removes the need for tiling functions (\texttt{tilize}/\texttt{untilize}), which account for $\approx$90\% of \textit{MatMul}'s CPU time.
Under UPM, the \textit{Axpy} method could match or exceed CPU baseline performance at a fraction of the energy consumption.
Even the \textit{MatMul} method becomes viable once the dominant conversion overhead is eliminated.

An additional benefit of APU-based architectures, not captured in these projections, is the reduction of accelerator initialization overhead.
Integrating CPU and accelerator within the same package minimizes the near-constant $\approx$1\,s startup cost observed in our experiments.

\subsection{Future Directions: Tenstorrent Blackhole}

Tenstorrent's next-generation Blackhole architecture~\cite{tt_blackhole_specs} addresses several of the limitations identified in this work and represents a significant step toward making RISC-V AI accelerators viable for HPC workloads.

The most transformative change is the inclusion of 16 SiFive x280 64-bit, dual-issue, in-order CPU cores arranged in four clusters~\cite{blackhole_hotchips, blackhole_microbench}.
These cores are sufficiently powerful to boot Linux and serve as an on-device host, fundamentally transforming Blackhole from a traditional PCIe accelerator requiring host supervision into a \emph{standalone AI computer}.
For our stencil workloads, this means that the scalar-heavy preprocessing (submatrix extraction, boundary handling, and format conversions) could be executed on the Big RISC-V cores directly on-chip, without PCIe round-trips.
The Big cores share 64\,MB of total L2 cache and have coherent access to 32\,GB of GDDR6 memory through a dedicated interconnect separate from the Tensix NoC~\cite{blackhole_microbench}.

In addition, Blackhole scales to 120 Tensix cores (up from 64 on Wormhole), with 1.5\,MB SRAM per core totaling 180\,MB of distributed on-chip memory~\cite{tt_blackhole_specs}.
This increased parallelism and local storage could better hide transfer latencies and reduce the need for off-chip memory access during stencil iterations.
It moves to PCIe Gen5 $\times$16, doubling available bandwidth compared to Wormhole's Gen4 interface. In addition, the transition from 12\,nm (Wormhole) to 6\,nm enables higher transistor density and improved power efficiency, supporting the increased core count within a 300\,W power envelope.
With Big RISC-V cores handling scalar preprocessing on-chip and 120 Tensix cores executing tiled computation, Blackhole approximates the UPM scenario analyzed in Section~\ref{sec:discussion}: data movement between the ``CPU'' and ``accelerator'' portions occurs over high-bandwidth on-chip interconnects rather than PCIe.
While Blackhole does not implement true UPM (the Big cores and Tensix cores access separate memory hierarchies), the tight integration within a single chip eliminates the dominant bottlenecks we identified.
Reassessing our methodologies on Blackhole is a clear and immediate direction for future work.

\vspace{-10px}
\section{Conclusions and Future Work}
\label{sec:conclusions}

We have presented a comprehensive analysis of mapping stencil computations onto the Tenstorrent Wormhole RISC-V AI accelerator through two heterogeneous CPU--accelerator methodologies.
The \textit{Axpy} method, which avoids tiled format conversions entirely, outperforms the \textit{MatMul} approach by 75$\times$ and demonstrates lower energy consumption than a CPU baseline despite being 3$\times$ slower.
Profiling reveals that the Wormhole kernel itself is competitive with CPU execution, since the performance gap is dominated by transfer overhead, device initialization, and CPU-side preprocessing.

Our theoretical analysis shows that unified memory architectures could close this gap: UVM reduces transfer overhead by 15$\times$, while UPM eliminates it entirely.
Tenstorrent's Blackhole architecture, with its 16 on-chip Big RISC-V cores capable of scalar computation and a standalone operating model, represents a concrete path toward realizing these benefits.

The most immediate direction for future work is evaluating our methodologies on Blackhole hardware, where the on-chip Big RISC-V cores could handle preprocessing without PCIe transfers. Beyond that, we plan to explore new algorithmic formulations to efficiently map stencils onto dataflow architectures, extend to multi-chip configurations leveraging Blackhole's Ethernet-based interconnect for distributed stencil computation, and investigate alternative PDE solvers to enable spectral-domain methods on Tenstorrent hardware.

\bibliographystyle{ACM-Reference-Format}
\bibliography{biblio}

@inproceedings{zeni2020logan,
  author    = {A. Zeni and G. Guidi and M. Ellis and N. Ding and M. D. Santambrogio and S. Hofmeyr and A. Bulu{\c{c}} and L. Oliker and K. Yelick},
  title     = {Logan: High-Performance {GPU}-Based X-Drop Long-Read Alignment},
  booktitle = {2020 IEEE International Parallel and Distributed Processing Symposium (IPDPS)},
  pages     = {462--471},
  publisher = {IEEE},
  year      = {2020}
}

@article{giannozzi2020quantum,
  author  = {P. Giannozzi and O. Baseggio and P. Bonf{\`a} and D. Brunato and R. Car and I. Carnimeo and C. Cavazzoni and S. {De Gironcoli} and P. Delugas and F. {Ferrari Ruffino} and others},
  title   = {Quantum {ESPRESSO} Toward the Exascale},
  journal = {The Journal of Chemical Physics},
  volume  = {152},
  number  = {15},
  year    = {2020}
}

@article{schmidhuber2015deep,
  author  = {J. Schmidhuber},
  title   = {Deep Learning in Neural Networks: An Overview},
  journal = {Neural Networks},
  volume  = {61},
  pages   = {85--117},
  year    = {2015}
}

@misc{tt_wormhole_specs,
  author       = {{Tenstorrent}},
  title        = {Wormhole Accelerator Specifications},
  year         = {2025},
  howpublished = {\url{https://docs.tenstorrent.com/aibs/wormhole/specifications.html}},
  note         = {Accessed: 2025-09-11}
}

@inproceedings{brown2024stencils,
  author    = {Nick Brown and Ryan Barton},
  title     = {Accelerating Stencils on the {Tenstorrent} {Grayskull} {RISC-V} Accelerator},
  booktitle = {SC24-W: Workshops of the International Conference for High Performance Computing, Networking, Storage and Analysis},
  pages     = {1690--1700},
  publisher = {IEEE},
  year      = {2024},
  doi       = {10.1109/SCW63240.2024.00211}
}

@misc{amd_mi300a,
  author       = {{AMD Corporation}},
  title        = {{AMD Instinct MI300A APU} Product Overview},
  year         = {2023},
  howpublished = {\url{https://www.amd.com/en/products/accelerators/instinct/mi300a.html}},
  note         = {Accessed: 2025-10-05}
}

@misc{nvidia_gh200,
  author       = {{NVIDIA Corporation}},
  title        = {{NVIDIA GH200 Grace Hopper Superchip}},
  year         = {2023},
  howpublished = {\url{https://www.nvidia.com/en-us/data-center/grace-hopper-superchip/}},
  note         = {Accessed: 2025-10-05}
}

@inproceedings{brown2025fft,
  author    = {Nick Brown and Jake Davies and Felix {Le Clair}},
  title     = {Exploring Fast {Fourier} Transforms on the {Tenstorrent} {Wormhole}},
  booktitle = {ISC 2025 Workshops},
  series    = {Lecture Notes in Computer Science},
  publisher = {Springer},
  year      = {2025},
  note      = {arXiv:2506.15437}
}

@inproceedings{amati2025nbody,
  author    = {Giorgio Amati and Matteo Turisini and Andrea Monterubbiano and Mattia Paladino and Elisabetta Boella and Daniele Gregori and Danilo Croce},
  title     = {Accelerating Gravitational {N}-Body Simulations Using the {RISC-V}-Based {Tenstorrent} {Wormhole}},
  booktitle = {Proceedings of the SC'25 Workshops of the International Conference for High Performance Computing, Networking, Storage and Analysis},
  publisher = {ACM},
  year      = {2025},
  doi       = {10.1145/3731599.3767528}
}

@inproceedings{chen2024convstencil,
  author    = {Yuetao Chen and Kun Li and Yuhao Wang and Donglin Bai and Lei Wang and Lingxiao Ma and Liang Yuan and Yunquan Zhang and Ting Cao and Mao Yang},
  title     = {{ConvStencil}: Transform Stencil Computation to Matrix Multiplication on Tensor Cores},
  booktitle = {Proceedings of the 29th ACM SIGPLAN Annual Symposium on Principles and Practice of Parallel Programming (PPoPP '24)},
  pages     = {333--347},
  publisher = {ACM},
  year      = {2024}
}

@article{cai2023assessing,
  author  = {Z. Cai and R. Giordano and others},
  title   = {Assessing {Tenstorrent}'s {RISC-V} Matrix Multiply Acceleration},
  journal = {arXiv preprint arXiv:2305.10314},
  year    = {2023}
}

@misc{blackhole_microbench,
  title        = {Dissecting the {Tenstorrent} {Blackhole} Architecture via Microbenchmarking},
  year         = {2025},
  howpublished = {\url{https://asplos.dev/wordpress/wp-content/uploads/2025/09/TT_bench-1.pdf}},
  note         = {Accessed: 2025-11-01}
}

@misc{tt_blackhole_specs,
  author       = {{Tenstorrent}},
  title        = {Blackhole Accelerator Specifications},
  year         = {2025},
  howpublished = {\url{https://docs.tenstorrent.com/aibs/blackhole/specifications.html}},
  note         = {Accessed: 2025-11-01}
}

@inproceedings{blackhole_hotchips,
  author    = {Jasmina Vasiljevic and Davor Capalija},
  title     = {{Blackhole} \& {TT-Metalium}: The Standalone {AI} Computer and Its Programming Model},
  booktitle = {Hot Chips 36},
  year      = {2024}
}

@misc{tracy,
  author       = {Bartosz Taudul},
  title        = {Tracy Profiler},
  howpublished = {\url{https://github.com/wolfpld/tracy}},
  year         = {2024}
}

@misc{Corsix2024,
  author       = {corsix},
  title        = {Community Highlight: Tenstorrent Wormhole Series Part 2: Which disabled rows?},
  howpublished = {Tenstorrent Newsroom},
  month        = nov,
  year         = {2024},
  url          = {https://tenstorrent.com/newsroom/community-highlight-tenstorrent-wormhole-series-part-2-which-disabled-rows},
  note         = {Accessed: 2024-11-18}
}

@online{TenstorrentTensixNeo,
  author       = {{Tenstorrent Inc.}},
  title        = {Tensix Neo: RISC-V-based AI IP For Extraordinary AI Performance},
  year         = {2024},
  url          = {https://tenstorrent.com/ip/tensix-neo},
  note         = {Accessed: 2024-11-18},
  organization = {Tenstorrent}
}

\end{document}